# Cloaking of Thermoelectric Transport

Troy Stedman and Lilia M. Woods*

Department of Physics, University of South Florida, Tampa FL 33620 USA

*Abstract*

The ability to control electromagnetic fields, heat currents, electric currents, and other physical phenomena by coordinate transformation methods has resulted in novel functionalities, such as cloaking, field rotations, and concentration effects. Transformation optics, as the underlying mathematical tool, has proven to be a versatile approach to achieve such unusual outcomes relying on materials with highly anisotropic and inhomogeneous properties. Most applications and designs thus far have been limited to functionalities within a single physical domain. Here we present transformation optics applied to thermoelectric phenomena, where thermal and electric flows are coupled via the Seebeck coefficient and Joule heating is taken into account. Using laminates, we describe a theoretical thermoelectric cloak capable of hiding objects from thermoelectric flow. Our calculations show that such a cloak does not depend on the particular boundary conditions and can also operate in different single domain regimes. These proof-of-principle results constitute a significant step forward towards finding unexplored ways to control and manipulate coupled transport.

**Introduction**

Designing metamaterials (MMs), artificial systems with functionalities unattainable with naturally occurring materials, is a promising direction with far reaching consequences[1-5]. Significant breakthroughs have been made in the field of optics, where man-made composites in which local magnetic dipole moments induced by oscillating electric currents can create strong magnetic response at optical frequencies[6]. Other optical MMs having negative magnetic permeabilities and giving rise to negative light phase velocities have also been made[2,3]. Optical devices, including invisibility cloaks, field concentrators, and rotators have been designed based on a medium with a spatially changing refractive index that alters light propagation pathways[7-13]. Some of this progress has relied on transformation optics (TO), a powerful mathematical tool that enables molding the electromagnetic energy flow in desired ways by using fictitious spatial distortions mapped into material composites capable of guiding light[14-17]. This idea combined with progress in fabrication techniques have enabled experimental realization of invisibility cloaks (including in the visible light range) and DC magnetic cloaks as well as new beam steering lenses, for example[18-21].

TO techniques have also been extended to other areas[4], where manipulation of heat fluxes[5,22-26], electric currents[27-30], mass diffusion, and acoustic propagation[31], have been demonstrated. The growing experimental evidence for heat flow cloaking, focusing, and reversal[4,5,24,32-34] has been especially useful to explore novel thermal phenomena and related applications. The success of TO beyond its original application in electromagnetic phenomena shows that this approach is quite general. We emphasize this point further by noting that in a closed system, heat and particle flows, for example, are governed by the laws of conservation of total energy and total number of particles with respective heat and particle diffusion equations. At the same time, the stationary continuity equation for the electric current, which takes into account Ohm's law, is mathematically equivalent to those for heat and particle flows. Therefore, although electromagnetic flow is associated with the wave-like nature of charge density propagation while heat and particle diffusion propagation are not mediated by waves, the physical behavior of this stationary transport is actually the same. This point has been explored recently by constructing bifunctional cloaks capable of guiding both, heat and electric currents[35-38]. Such single devices can be especially useful in applications where the simultaneous control of thermal and electrical phenomena is needed, like in solar cells and thermoelectric devices.

TO techniques have been indispensable in the design of MMs for bifunctional cloaks, which can operate in the simultaneous presence of temperature and voltage gradients. Even though such composites constitute an important step forward to access a multiphysics domain of operation, the neglect of thermoelectric (TE) phenomena, which characterize the coupling of heat and electrical transport, is a major drawback. The thermodynamic description of this coupled transport includes Joule heating and is captured by the Seebeck coefficient $S$, a material property describing the production of a voltage drop due to an applied temperature difference. This property reflects the fact that charge carriers transport heat and electricity simultaneously, a manifestation of the charge carrier specific heat[39]. Therefore, coupled heat and electric transport in materials in which the Seebeck coefficient is appreciable cannot be manipulated and controlled by the bifunctional cloaks described in[35-38].

In this work the following fundamental question is addressed: Can the versatility of TO techniques be used to mold thermoelectric flow and achieve effects (such as cloaking, for example) not possible to observe with natural materials? This problem requires generalizing the diffusive nature of thermodynamic flows by taking into account the coupling between heat and electricity via the Seebeck coefficient. Here, we show that this is possible and use the results to obtain a TE cloak capable of hiding objects without disturbing the coupled external heat and electric currents. Such a TE cloak operates under *any* thermal and electrical boundary conditions, generalizing the bifunctional cloak. This is achieved by using the form invariance of the governing laws of energy and charge conservation under coordinate transformations as required by the TO method. Similar to other types of cloaks, TE cloaking demands materials with highly anisotropic and inhomogeneous properties. Since such materials are not readily available in nature, using bilayer composites we construct laminate MMs by finding the specific properties necessary to achieve a cloaking effect.

**Results**

**Basic Equations:** The direct conversion of heat into electricity and vice versa constitutes the TE effect, which reflects the production of a charge current flow from a heat current and heat flow due to a voltage difference[40]. These two reciprocal phenomena are the Peltier and Seebeck effects respectively. In the former, a heat current occurs due to isothermal current flow, while in the latter an electric current is generated due to a temperature difference. A unified representation of TE phenomena based on governing conservation laws and linear constitutive relations conforming to the general principles of thermodynamics is given by the Onsager-de Groot-Callen theory[41,42]. In a steady-state with local equilibrium, TE transport is described by taking into consideration the electrochemical potential $\mu = \mu_C + eV$ ($\mu_C$ – chemical potential, $V$ – electric potential) and temperature $T$ of the system with governing equations,

$$\vec{\nabla} \cdot \boldsymbol{J} = 0 \ , \ \vec{\nabla} \cdot \boldsymbol{J_Q} = -\vec{\nabla}\mu \cdot \boldsymbol{J}, \tag{1}$$

where $\boldsymbol{J}$ is the electric current density and $\boldsymbol{J_Q}$ is the heat current density. The first equation expresses local steady-state charge conservation, while the second one corresponds to local steady-state energy conservation whose right hand side is the flux due to Joule heating. The constitutive equations are the linear relations from the Onsager-de Groot-Callen theory between fluxes and driving forces due to gradient fields. The TE driving forces are the electrochemical potential and temperature gradients while the constitutive equations are

$$\boldsymbol{J} = -\overleftrightarrow{\sigma} \cdot \vec{\nabla}\mu - \overleftrightarrow{\sigma} \cdot \overleftrightarrow{S} \cdot \vec{\nabla}T \ , \ \boldsymbol{J_Q} = -\overleftrightarrow{\kappa} \cdot \vec{\nabla}T - T\overleftrightarrow{S}^t \cdot \overleftrightarrow{\sigma} \cdot \overleftrightarrow{S} \cdot \vec{\nabla}T - T\overleftrightarrow{S}^t \cdot \overleftrightarrow{\sigma} \cdot \vec{\nabla}\mu, \tag{2}$$

where the electrical conductivity, $\overleftrightarrow{\sigma}$, the thermal conductivity, $\overleftrightarrow{\kappa}$, and the Seebeck coefficient, $\overleftrightarrow{S}$, are Cartesian tensors. The Onsager reciprocal requirement that the electrical and thermal conductivities equal their transpose counterparts, $\overleftrightarrow{\sigma} = \overleftrightarrow{\sigma}^t$ and $\overleftrightarrow{\kappa} = \overleftrightarrow{\kappa}^t$, is accounted for in equation (2). One notes that Ohm's law and Fourier's law are the first terms in $\boldsymbol{J}$ and $\boldsymbol{J}_Q$, respectively, and describe the independent production of charge and heat currents under their corresponding gradients. The remaining terms are the TE effects reflecting the coupled charge-heat transport, such that a temperature gradient can lead to a charge carrier flux and that the charge carriers can transport heat flux as well.

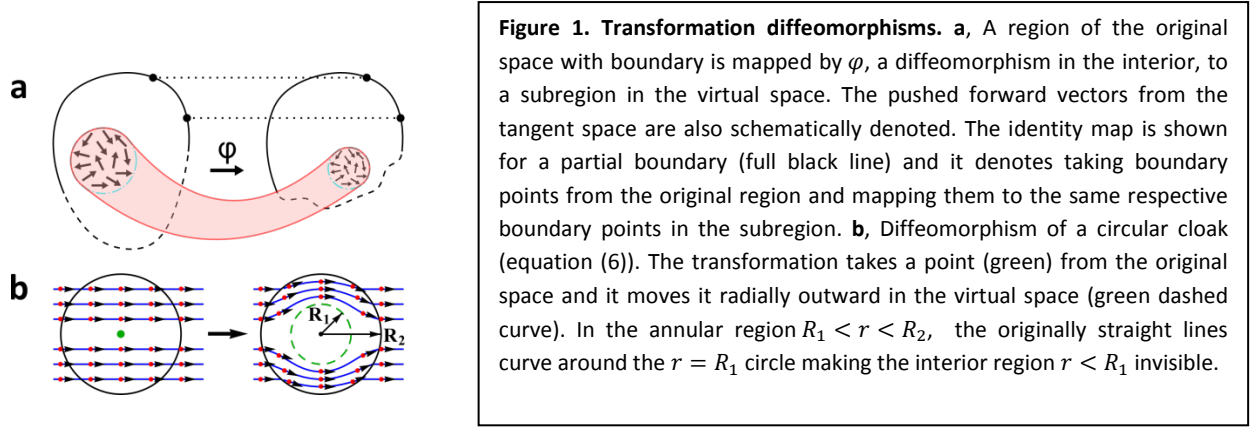

**Figure 1. Transformation diffeomorphisms. a**, A region of the original space with boundary is mapped by $\varphi$, a diffeomorphism in the interior, to a subregion in the virtual space. The pushed forward vectors from the tangent space are also schematically denoted. The identity map is shown for a partial boundary (full black line) and it denotes taking boundary points from the original region and mapping them to the same respective boundary points in the subregion. **b**, Diffeomorphism of a circular cloak (equation (6)). The transformation takes a point (green) from the original space and it moves it radially outward in the virtual space (green dashed curve). In the annular region $R_1 < r < R_2$, the originally straight lines curve around the $r = R_1$ circle making the interior region $r < R_1$ invisible.

**Transformation Optics for Thermoelectricity:** In this paper we demonstrate cloaking of TE transport, where a region is invisible from both heat and charge fluxes *simultaneously* in the presence of thermal and electrochemical potential gradients while currents and gradients external to the cloak are unaltered. For this purpose, we utilize TO techniques which explores the form invariance of the underlying equations and maps fictitious distortions into spatially inhomogeneous and anisotropic material properties. These concepts have been effectively extended to find appropriate spatial variations of the material properties to achieve other effects including flow rotators, inverters, and concentrators in electrodynamics, heat conduction, and acoustics, for example. We show that TO techniques can be applied to thermoelectricity by relying on the central idea that the governing equations (equations (1),(2)) are invariant under coordinate transformations. Thus the TE fluxes can be modified in a prescribed way (such as cloaking) by making a suitable choice for a particular coordinate transformation $\boldsymbol{r}' = \boldsymbol{r}'(\boldsymbol{r})$.

The underlying mathematical foundation concerns a region with a boundary in a given medium, commonly referred to as "original space", where materials properties (tensors, in general) along with relevant vector fields, such as charge and heat fluxes, are specified. The vector fields in the region are pushed forward via a smooth mapping based on a diffeomorphism, $\varphi$, to some subregion, commonly referred to as "virtual space". This is schematically shown in Fig. 1a where the original space has a boundary. We note that $\varphi$ must be the identity mapping on the boundary, a necessary requirement for invisibility cloaking, which ensures that the boundary points in the original space are transformed to the same respective boundary points in the virtual space[17]. The essence of TO is that the pushforward map is such that the governing and constitutive equations are not only satisfied in the subregion but are also form invariant. The manipulations required by TO consist of spatial stretching and/or compressing but no spatial tearing to create a new vector field by the pushforward map. To achieve this physically, the materials properties in the subregion must be changed. Figure 1b illustrates a particular circular cloak mapping where the center of a circle is blown up to a larger circle. This distortion results in modified material properties which become highly anisotropic and inhomogeneous. Importantly, integral curves (curves that are tangent to the vector field at every point on the curve) under a global diffeomorphism remain integral curves of the vector fields pushed forward by a diffeomorphism, giving a nice picture of the image of integral curves. Simply put, the vector fields "follow" the distortion as shown in Fig. 1b where straight lines now curve around the cloaked region.

The TO techniques described above are now applied to thermoelectricity. Vectors and material properties are pushed forward via a diffeomorphism applied to a region with an identity map on the boundary, while the governing and constitutive equations must be invariant under such a transformation. This is achieved if the transformed properties (primed) are related to the original ones (unprimed) as

$$\overleftrightarrow{\sigma}' = \frac{\overleftrightarrow{A} \cdot \overleftrightarrow{\sigma} \cdot \overleftrightarrow{A}^t}{|\overleftrightarrow{A}|} \quad , \quad \overleftrightarrow{\kappa}' = \frac{\overleftrightarrow{A} \cdot \overleftrightarrow{\kappa} \cdot \overleftrightarrow{A}^t}{|\overleftrightarrow{A}|} \quad , \quad \overleftrightarrow{S}' = \overleftrightarrow{A}^{-t} \cdot \overleftrightarrow{S} \cdot \overleftrightarrow{A}^t, \tag{3}$$

where $\overleftrightarrow{A}$ is the Jacobian matrix with elements $A^i_j = \frac{\partial x'^i}{\partial x^j}$ and the transformed coordinates $\boldsymbol{r}' = (x'^1, x'^2, x'^3)$ are related to the original ones $\boldsymbol{r} = (x^1, x^2, x^3)$ by a smooth, invertible function $\boldsymbol{r}'(\boldsymbol{r})$. $\overleftrightarrow{A}^t$ is the matrix transpose of $\overleftrightarrow{A}$, $\overleftrightarrow{A}^{-t}$ is the inverse of $\overleftrightarrow{A}^t$, and $|\overleftrightarrow{A}| = \text{Det } \overleftrightarrow{A}$. It is important to note that for a homogeneous and isotropic medium, this transformation gives $\overleftrightarrow{S}' = \overleftrightarrow{S} = S$, which shows that the Seebeck coefficient in this region is unaltered by the transformation. The electric and heat current vector fields together with the electrochemical potential and temperature gradients are pushed forward as

$$J' = \frac{\tilde{A}}{|\tilde{A}|} \cdot J \ , \ J'_Q = \frac{\tilde{A}}{|\tilde{A}|} \cdot J_Q, \tag{4}$$

$$\vec{\nabla}\mu' = \vec{A}^{-t} \cdot \vec{\nabla}\mu \ , \ \vec{\nabla}T' = \vec{A}^{-t} \cdot \vec{\nabla}T. \tag{5}$$

**A Circular Thermoelectric Cloak:** The relations in equations (3)-(5) show that a given coordinate transformation $r'(r)$ preserves the form invariance of the governing and constitutive equations, while the physical properties and currents change accordingly. Thus to achieve effects, such as cloaking, focusing, or reversal of TE transport, one must specify an appropriate coordinate transformation, which will generally result in anisotropic and inhomogeneous material properties according to equation (3). Here we consider the case of a 2D circular TE cloak, although this TO approach can be generalized to 3D as well. The circular cloak, consisting of an $R_1 < r < R_2$ annular region, can hide an object placed within the interior $r < R_1$ from any heat or electric currents in the external medium $r > R_2$ (Fig. 1b). The corresponding diffeomorphism compresses points in the interior of the outer circle of radius $R_2$ into the annular region and is given by

$$r' = \left(\frac{R_2 - R_1}{R_1}\right)r + R_1 \ , \ \theta' = \theta. \tag{6}$$

In effect, field lines that enter the cloak are maneuvered around the inner circle with radius $R_1$ and exit the outer circle with radius $R_2$ at the same point that they would if there were no cloak at all. The fields on the outer circle also have the same amplitude that they would if there were no cloak at all. Therefore, the cloaking effect isolates the interior region from any currents and gradients external to the cloak while leaving the currents and gradients in the external medium unaffected by the cloak. Applying this transformation to an isotropic and homogeneous medium with scalar electrical conductivity $\sigma$, thermal conductivity $\kappa$, and Seebeck coefficient $S$, we find

$$\overleftrightarrow{\sigma}' = \sigma R(\theta) T(r) R^t(\theta); \ \overleftrightarrow{\kappa}' = \kappa R(\theta) T(r) R^t(\theta); \ S' = S \tag{7}$$

where,

$$R(\theta) = \begin{pmatrix} \cos\theta & -\sin\theta \\ \sin\theta & \cos\theta \end{pmatrix}, \quad T(r) = \begin{pmatrix} \frac{r - R_1}{r} & 0 \\ 0 & \frac{r}{r - R_1} \end{pmatrix}. \tag{8}$$

Let us note here that $(r, \theta)$ specify the coordinates of the cloak and are the codomain coordinates under the considered diffeomorphism (Eq. 6) and hence are different from the unprimed domain coordinates. Clearly the electrical and thermal conductivities of the cloak (equation (7)) are anisotropic and inhomogeneous. The Seebeck coefficient, on the other hand, is simply that of the medium. To see the cloaking effect due to this transformation, we consider the radial and azimuthal conductivities $\sigma'_{rr} = \sigma \frac{r - R_1}{r}$ and $\sigma'_{\theta\theta} = \sigma \frac{r}{r - R_1}$, respectively. One notes that $\sigma'_{rr} \to 0$, while $\sigma'_{\theta\theta} \to \infty$ as $r \to R_1$ so that the cloak only has an azimuthal electrical response and no radial electrical response near the outer edge of the inner circle. The radial and azimuthal thermal conductivities have the same behavior as their electrical counterparts. Consequently, the electric and heat currents comprising the TE flow are guided around the inner circle.

Figure 2a shows our simulation results for the TE cloak from equations (6)-(8) obtained using the finite element based COMSOL MULTIPHYSICS package with TE boundary conditions (as described in Methods). It is evident that due to $\kappa'_{rr} \to 0, \sigma'_{rr} \to 0$ as $r \to R_1$ no heat and electric currents penetrate the $r < R_1$ cloaked region. After reaching a steady state, a constant temperature and constant potential profile in the cloaked region are achieved. The simulations show that the constant temperature and constant potential inside the cloaked region are $\frac{T_1+T_2}{2}$ and $\frac{V_1+V_2}{2}$, respectively ($T_1, V_1$ – temperature and potential at the left end of the medium; $T_2, V_2$- temperature and potential at the right end of the medium). The same constant temperature behavior has also been found for thermal cloaks under similar simulations conditions[22]. Figure 2a further shows that the heat and electric currents as well as the temperature and potential gradients outside of the cloaking region, $r > R_2$, are the same as those in the isotropic medium if there were no cloak, indicating that these currents and gradients are unperturbed by the presence of the cloak. As a result, an object in the $r < R_1$ region will not experience any effects from the TE flow outside of this region, while the $r > R_2$ region will be insensitive to any TE flow from the interior region.

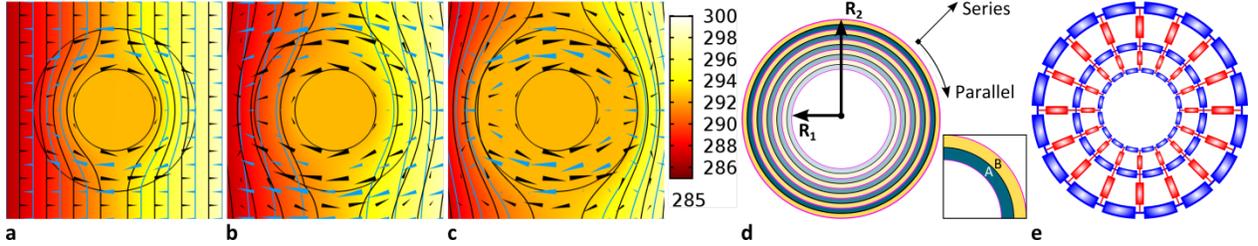

**Figure 2. Cloak simulations and schematics. a**, An ideal TE cloak with $\sigma'_{rr} = \sigma \frac{r-R_1}{r}$; $\sigma'_{\theta\theta} = \sigma \frac{r}{r-R_1}$; $\kappa'_{rr} = \kappa \frac{r-R_1}{r}$; $\kappa'_{\theta\theta} = \kappa \frac{r}{r-R_1}$; $S' = S$ for the cloak material properties for $R_1 < r \leq R_2$ with "TE" boundary conditions for the simulations. The heat and electric currents are shown in blue and black cones, respectively. The blue and black curves are the isotherms and equipotentials, respectively. The background color scheme indicates the temperature profile. Although not shown, the voltage profile has a similar behavior as the temperature profile. **b**, An otherwise ideal TE cloak with $S' = 0.1 * S$. **c**, An otherwise ideal TE cloak with $S' = 10 * S$. **d**, Concentric bilayers for the circular TE cloak of inner radius $R_1$ and outer radius $R_2$. The bilayers have equal thickness and are made up of two different homogeneous and isotropic materials, denoted as A and B. **e**, Schematic representation of the electrical conductivity of the layered cloak in terms of a configuration of resistors that are radially in series (red) and azimuthally in parallel (blue) (also denoted in **d**). Relevant details for the simulations and boundary conditions are shown in Methods.

Let us further note that $\overleftrightarrow{\kappa}'$ from equation (3) is the same as the requirement for the thermal cloak investigated previously[22]. Therefore, it is concluded that the TE cloak here will operate also as a thermal cloak under an applied temperature difference only. The same conclusion can be reached for an electric cloak when only an electric potential difference is applied[27]. Furthermore, the TE cloak will also operate as a bifunctional cloak, as described in[45], in the presence of simultaneous temperature and electric potential differences when the medium has negligible Seebeck coefficient. However, if the medium has an appreciable $S$ and the $R_1 < r < R_2$ region does not have the Seebeck coefficient according to equation (7), there is a significant coupling between the thermal and electric flows and the cloak cannot operate as the bifunctional cloak. To illustrate this point further, in Fig. 2b,c we show simulations for a cloak with $\overleftrightarrow{\kappa}', \overleftrightarrow{\sigma}'$ transforming according to equations (7),(8) and $S' = 0.1 * S$ and $S' = 10 * S$ respectively under the TE boundary conditions. One finds that the isotherms and equipotentials are distorted near the outer edge of the cloak and the heat and electric currents differ from those for the ideal cloak in Figure 2a. The distortions of the currents appear to be greater near the top and bottom edges outside of the cloak for the $S' = 10*S$ case, while distortions of the currents for the cloak with $S' = 0.1*S$ are especially different near the left and right edges outside of the cloak as compared to Fig. 2a. Also, the electric current in the cloak with $S' = 0.1*S$ has reversed direction from the one of the ideal cloak. In effect, the TE transport in Fig. 2b,c consisting of the electric and heat currents is not cloaked.

**Laminate MMs for TE cloaking:** As pointed out earlier, the ideal TE cloak in Fig. 2a requires not only anisotropic and inhomogeneous properties, but it requires having $\vec{\sigma}'$ and $\vec{\kappa}'$ components that grow without bound as $r \to R_1$, making it impossible to realize such a behavior physically, let alone with naturally occurring materials. Therefore, we appeal to MMs to create a cloak with effective material properties that approximate their ideal pushed forward values (equation (7)). To this end, we consider a composite consisting of concentric bilayers, such that each one is composed of two isotropic and homogeneous layers of equal thickness (denoted as A and B in Fig. 2d), a strategy also used in[9,22-24,43,44]. If the thickness of the bilayers is small compared to the inner radius $R_1$, we can approximate the two layers $A$ and $B$ as radially in series and azimuthally in parallel[4,9,44] (as schematically shown in Fig. 2e), giving rise to effective material properties according to

$$\sigma'_{rr} = 2\frac{\sigma_A \sigma_B}{\sigma_A + \sigma_B}, \quad \sigma'_{\theta\theta} = \frac{\sigma_A + \sigma_B}{2}, \quad \kappa'_{rr} = 2\frac{\kappa_A \kappa_B}{\kappa_A + \kappa_B}, \quad \kappa'_{\theta\theta} = \frac{\kappa_A + \kappa_B}{2} \tag{9}$$

$$\sigma'_{rr} S'_{rr} = 2\frac{\sigma_A S_A \sigma_B S_B}{\sigma_A S_A + \sigma_B S_B}, \quad \sigma'_{\theta\theta} S'_{\theta\theta} = \frac{\sigma_A S_A + \sigma_B S_B}{2}. \tag{10}$$

Thus we arrive at the following bilayer materials properties:

$$\sigma_{A(B)} = \sigma D_{\pm(\mp)}, \kappa_{A(B)} = \kappa D_{\pm(\mp)}, \tag{11}$$

$$S_{A(B)} = S \quad \text{or} \quad S_{A(B)} = S\frac{D_{\pm(\mp)}}{D_{\mp(\pm)}}, \tag{12}$$

The $D_\pm = \frac{r}{r-R_1}\left(1 \pm \sqrt{1-\left(\frac{r-R_1}{r}\right)^2}\right)$ factor determines the anisotropic nature of the bilayer properties. Noting that $D_+ D_- = 1$, one finds that if $\sigma_A = \sigma D_+$ then $\sigma_B = \sigma/D_+$. Similar proportionality relations are obtained for $\kappa_{A(B)}$. Interestingly there are two choices for the bilayer Seebeck coefficients. In one case, $S$ must be the same for both layers (thus it must be the same throughout the entire cloak), while in the second case $S$ is determined by the $D_\pm$ ratio such that if $S_A = SD_+^2$ then $S_B = S/D_+^2$. Any combination of choices (all possibilities are shown in Methods) for the layer material properties in equations (11),(12) satisfies the bilayer approximation of the ideal pushed forward material properties in equation (7), and the $\pm \to \mp$ correspond to simply swapping the A and B layers. In Methods we give explicitly all possible combinations of layer properties, which further shows that the requirements in equations (11),(12) are independent of each other.

The bilayer building blocks can now be used to construct a MM laminate circular cloak by using the guidelines found in equations (9)-(12). The laminate is taken to be composed of $N$ concentric bilayers of equal thickness $2d$ with $d = \frac{R_2-R_1}{2N}$ being the thickness of the $A_n$ and $B_n$ layers comprising the $n^{th}$ bilayer. The properties are $\sigma_{A_n} = \sigma_-\big((2n-1)d\big)$, $\kappa_{A_n} = \kappa_-\big((2n-1)d\big)$ and $\sigma_{B_n} = \sigma_+(2nd)$, $\kappa_{B_n} = \kappa_+(2nd)$ while $S_{A_n} = S_{B_n} = S$ with $n = 1,2,\ldots,N$, corresponding to the bilayer choice $\{\sigma_-, \sigma_+\}$, $\{\kappa_-, \kappa_+\}$, $\{S,S\}$ in Table I. This composite has layers with largely varying thermal and electrical conductivities, while the Seebeck coefficient stays the same. Here we also illustrate how the laminate TE cloak functions under different boundary conditions. In addition to results for the steady state simulations for $N = 1, 5,$ and $10$ under TE boundary conditions, shown in Fig. 3a-c, results for the same systems under transverse boundary conditions (described in Methods) are shown in Fig. 3d-f. The overall comparison of these results shows that the cloaking effect is present regardless of the boundary conditions and the performance improves as $N$ increases. Under TE boundary conditions for $N = 1$ (Fig. 3a) some isotherms penetrate the cloaked $r < R_1$ region, while the isotherms and equipotentials outside the cloak, $r > R_2$, close to the laminate are distorted. This means that the cloaking effect is imperfect and the medium itself has detectable isotherm and equipotential distortions. Also, the temperature and voltage distributions in $r < R_1$ is not uniform, as is the case of an ideal cloak shown in Fig. 2a. Under transverse boundary conditions, for $N = 1$ (Fig. 3d) some isotherms penetrate the cloaked region $r < R_1$, but the pattern for the isotherms and equipotentials is different as compared to Fig. 3a. These unwanted effects for both sets of boundary conditions diminish as $N$ increases due to decreasing of the thickness of the $A$ and $B$ layers which improves the approximation in equations (9),(10).

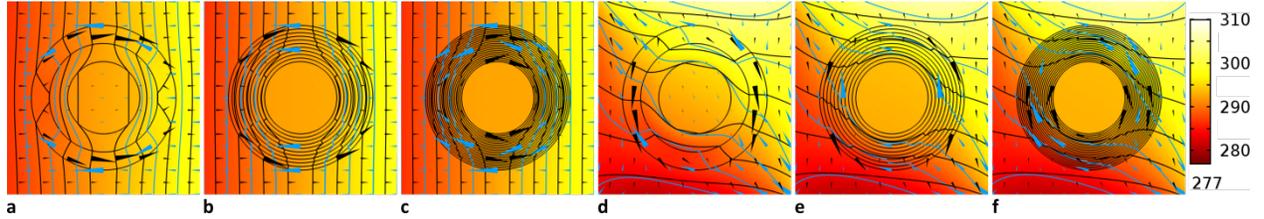

**Figure 3. Layered cloaks.** Simulations are shown of a $N$-bilayered laminate TE cloak with TE boundary conditions for **a**, $N = 1$, **b**, $N = 5$, **c**, $N = 10$. Simulations are shown of a $N$-bilayered laminate TE cloak with transverse boundary conditions for **d**, $N = 1$, **e**, $N = 5$, **f**, $N = 10$. The blue cones and curves correspond to the heat current and isotherms, respectively, while the black cones and curves correspond to the electric current and equipotentials, respectively. The background color scheme indicates the temperature profile.

To further examine the quality of the TE cloak, results for the temperature and voltage profiles along a horizontal line passing through the center of medium are shown for several cases in Fig. 4, including an ideal cloak, layered cloaks with $N = 1, 3,$ and $10$ bilayers and no cloak present for the two sets of boundary conditions. The temperature and potential for the ideal cloak are constant, such that $T = \frac{T_1+T_2}{2}$ and $V = \frac{V_1+V_2}{2}$. Figure 4 also shows that the largest deviations are found for a laminate composite with $N = 1$ bilayer. As $N$ increases, these $T$ and $V$ vs $x$ profiles approach the ideal behavior, indicating improved cloaking performance. The TE cloaking effect can also be quantified by evaluating the standard temperature and potential deviations of the laminate composites, defined respectively as $\bar{\sigma}_{STD,T} = \sqrt{\frac{1}{M}\sum_{i=1}^{}(\Delta T_i^c - \overline{\Delta T^c})^2}$ and $\bar{\sigma}_{STD,V} = \sqrt{\frac{1}{M}\sum_{i=1}^{}(\Delta V_i^c - \overline{\Delta V^c})^2}$. These are calculated on $M$ grid points for $r > R_2$, such that $\Delta T_i^c = T_i^c - T_i^m$ ($\Delta V_i^c = V_i^c - V_i^m$), $\overline{\Delta T^c} = \frac{1}{M}\sum_i \Delta T_i^c$ $\left(\overline{\Delta V^c} = \frac{1}{M}\sum_i \Delta V_i^c\right)$ where $T_i^c$ ($V_i^c$) is the temperature (voltage) with the cloak and $T_i^m$ ($V_i^m$) is the temperature (voltage) without the cloak at the $i$-th grid point[45]. As expected, the results for $\bar{\sigma}_{STD,T}$ and $\bar{\sigma}_{STD,V}$ (shown as inserts in Fig. 4), indicate that increasing $N$ results in a better cloaking effect approaching the $\bar{\sigma}_{STD,T} = \bar{\sigma}_{STD,V} = 0$ for the ideal cloak.

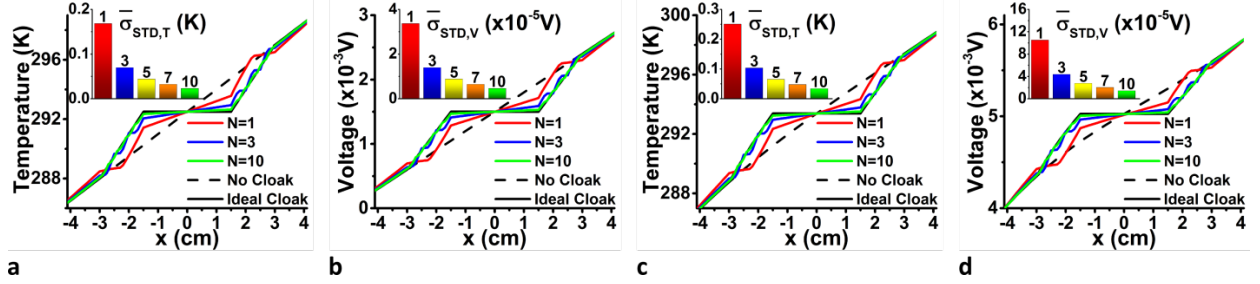

**Figure 4. Quantification of the cloaks' efficacies.** Characteristic profiles as a function of horizontal position $x$ for a horizontal line through the center of the medium is shown for simulations for **a**, temperature, **b**, voltage under TE boundary conditions, and **c**, temperature, **d**, voltage under transverse boundary conditions for a medium with no cloak, an ideal cloak, and laminate cloaks with $N = 1, 3$, and $10$ bilayers. The corresponding standard deviations between a medium with laminate cloaks with $N = 1, 3, 5, 7$, and $10$ bilayers and a medium with no cloak, calculated on a grid with M=100X100 points, are shown as inserts in each panel.

With these results, we see that the laminate cloaks improve as the number of bilayers increases, at the same time, the cloaking effect is unaffected by the boundary conditions. This is because the temperature and voltage profiles along with the distribution of heat and electric currents in the $r > R_2$ region do not affect the cloaking, emphasizing the generality of the TE cloak obtained via TO. This is in contrast with previous reports of cloaking or other effects obtained via scattering methods, which are strongly dependent on the specific boundary conditions[36,37].

We further discuss the material characteristics for a practical realization of TE cloaking. The laminate composite, as shown above, would have largely different electrical and thermal conductivities for the two layers in each bilayer with the largest variation in the innermost layers of the cloak. Although this is similar to the situation of thermal[24] and dc electric cloaks[27], an additional difficulty for the TE laminates arises from the requirements imposed for the Seebeck coefficient. We find that for the $N = 10$ bilayer case (Fig. 3 and Fig. 4), for example, $\sigma_{A_1} \sim \sigma/40, \sigma_{B_1} \sim 40\sigma, \kappa_{A_1} \sim \kappa/40, \kappa_{B_1} \sim 40\kappa$, and $\sigma_{A_{10}} \sim \sigma/4, \sigma_{B_{10}} \sim 4\sigma, \kappa_{A_{10}} \sim \kappa/4, \kappa_{B_{10}} \sim 4\kappa$. Thus while the electrical and thermal conductivities potentially may encompass values typical for metals and insulators, $S$ is constant throughout. Taking the $N = 10$ case with $S_A = SD_+^2$, $S_B = S/D_+^2$ (columns four, five, and six in the fourth row of Table 1 in Methods), we find that $S_{A_1} \sim 1800S, S_{B_1} \sim S/1800, S_{A_{10}} \sim 16S, S_{B_{10}} \sim S/16$, while $\sigma_{A_1} \sim \sigma/40, \sigma_{B_1} \sim 40\sigma, \kappa_{A_1} \sim \kappa/40, \kappa_{B_1} \sim 40\kappa$, and $\sigma_{A_{10}} \sim \sigma/4, \sigma_{B_{10}} \sim 4\sigma, \kappa_{A_{10}} \sim \kappa/4, \kappa_{B_{10}} \sim 4\kappa$. For a cloak with $N = 5$ the numerical factors are reduced by 2. There is a significant variation in $S$, which may be difficult to satisfy in practice, while the variation in the electrical and thermal conductivities is much less. We suggest that there are several routes for optimization in finding suitable materials for practical realization of a TE cloak. In addition to the several choices of bilayer combinations essentially arising from the properties being independent of each other, one might consider using a cloak with fewer layers. Even though the cloaking effect may be reduced, the variation in the materials properties is also reduced. Also, the location where $D_\pm(r)$ are evaluated, the radius and thickness of each layer, and materials doping can be further explored. Clearly this direction needs further investigation from a materials point of view.

**Discussion**

Here we have shown that electric and thermal transport coupled via thermoelectric phenomena can be manipulated according to virtual spatial distortions. Utilizing the form invariance of the underlying equations under coordinate transformations, the desired distortions are mapped into anisotropic and inhomogeneous thermoelectric properties of the materials. These ideas are applied to thermoelectric cloaking, which constitutes a significant step forward towards accessing multiphysics transformations with coupling between the domains. Further benefits can be drawn by realizing that the designed layered multifunctional metamaterials can operate as thermal or electric cloaks and are independent of the boundary conditions of operation. The several options for the materials and dependence on a variety of geometrical factors give pathways for property optimizations making TE cloaks practically possible.

**Methods**

Designing the laminate MM for TE cloaking shows that there are several choices for the transport characteristics of the individual layers, according to equations (11),(12). The possible combinations of the bilayer properties are given in Table I.

| $\{\sigma_A, \sigma_B\}$ | $\{\kappa_A, \kappa_B\}$ | $\{S_A, S_B\}$ | $\{\sigma_A, \sigma_B\}$ | $\{\kappa_A, \kappa_B\}$ | $\{S_A, S_B\}$ | $\{\sigma_A, \sigma_B\}$ | $\{\kappa_A, \kappa_B\}$ | $\{S_A, S_B\}$ |
|---|---|---|---|---|---|---|---|---|
| $\{\sigma_+, \sigma_-\}$ | $\{\kappa_+, \kappa_-\}$ | $\{S, S\}$ | $\{\sigma_+, \sigma_-\}$ | $\{\kappa_+, \kappa_-\}$ | $\left\{S\frac{D_+}{D_-}, S\frac{D_-}{D_+}\right\}$ | $\{\sigma_+, \sigma_-\}$ | $\{\kappa_+, \kappa_-\}$ | $\left\{S\frac{D_-}{D_+}, S\frac{D_+}{D_-}\right\}$ |
| $\{\sigma_+, \sigma_-\}$ | $\{\kappa_-, \kappa_+\}$ | $\{S, S\}$ | $\{\sigma_+, \sigma_-\}$ | $\{\kappa_-, \kappa_+\}$ | $\left\{S\frac{D_+}{D_-}, S\frac{D_-}{D_+}\right\}$ | $\{\sigma_+, \sigma_-\}$ | $\{\kappa_-, \kappa_+\}$ | $\left\{S\frac{D_-}{D_+}, S\frac{D_+}{D_-}\right\}$ |
| $\{\sigma_-, \sigma_+\}$ | $\{\kappa_+, \kappa_-\}$ | $\{S, S\}$ | $\{\sigma_-, \sigma_+\}$ | $\{\kappa_+, \kappa_-\}$ | $\left\{S\frac{D_+}{D_-}, S\frac{D_-}{D_+}\right\}$ | $\{\sigma_-, \sigma_+\}$ | $\{\kappa_+, \kappa_-\}$ | $\left\{S\frac{D_-}{D_+}, S\frac{D_+}{D_-}\right\}$ |
| $\{\sigma_-, \sigma_+\}$ | $\{\kappa_-, \kappa_+\}$ | $\{S, S\}$ | $\{\sigma_-, \sigma_+\}$ | $\{\kappa_-, \kappa_+\}$ | $\left\{S\frac{D_+}{D_-}, S\frac{D_-}{D_+}\right\}$ | $\{\sigma_-, \sigma_+\}$ | $\{\kappa_-, \kappa_+\}$ | $\left\{S\frac{D_-}{D_+}, S\frac{D_+}{D_-}\right\}$ |

**Table I.** A list of all possible choices for the material properties of each pair of layers comprising the bilayer laminate cloak.

The steady state simulations of equations (1),(2) are performed using finite element analysis in the COMSOL MULTIPHYSICS package. These governing and constitutive thermoelectric equations are implemented through the built in "Electric Currents" interface and a "Coefficient Form PDE" interface tailored to such equations. The annular cloak region (Fig. 2 and Fig. 3) is taken to be centered in a 10cm x 10cm material with $R_1 = 1.5$ cm and $R_2 = 3$ cm. The isotropic and homogeneous material is taken to have $\sigma = 3000 \frac{S}{cm}$; $\kappa = 1 \frac{W}{m*K}$; $S = -200 \frac{\mu V}{K}$, which are similar to those of the common TE material Bi$_2$Te$_3$[40].

The TE boundary conditions in Fig. 2a-c and Fig. 3a-c include electrically and thermally insulated top and bottom ends, which is commonly used in TE devices. Also, the left end is electrically grounded and held at a temperature $T_1 = 285$ K and the right end has an outward normal current density $J_n = 1 \frac{\text{A}}{\text{m}^2}$ and held at a temperature $T_2 = 300$ K. The chemical potential $\mu_C$ is constant throughout the entire material.

The transverse boundary conditions in Fig. 3d-f include electrically insulated left and right ends while the top and bottom ends are thermally insulated. Also, the temperature of the left end is held at $T_1 = 285$ K, the temperature of the right end is held at $T_2 = 300$ K, the bottom end is grounded and, the potential of the top end is held at $V = 0.01$ V.

**Data availability:** The data that support these findings are available from the corresponding author upon request.

**Acknowledgements**

We acknowledge financial support from the US National Science Foundation under Grant No. DMR-1400957. The use of the University of South Florida Research Computing facilities are also acknowledged. Discussions with Dr. Timothy Fawcett and Prof. David Rabson are also acknowledged.

**Author contributions**

All authors contributed equally to this work.

**Additional Information**

Competing Financial Interests**:** The authors declare no competing financial interests.

*Correspondence: lmwoods@usf.edu